\documentclass{article}

\usepackage{placeins}
\usepackage{booktabs}
\usepackage{graphicx}
\usepackage{wrapfig}
\usepackage{amsmath}
\usepackage{float}
\usepackage{url}

\usepackage[hidelinks]{hyperref}
\newcommand*{\email}[1]{\href{mailto:#1}{\nolinkurl{#1}} }

\usepackage[a4paper, margin=1in]{geometry}

\usepackage{tikz}
\usetikzlibrary{snakes}
\usetikzlibrary{positioning}
\usetikzlibrary{arrows.meta}
\usetikzlibrary{calc}
\counterwithin{figure}{section}
\counterwithin{table}{section}

\usepackage{lipsum}
\usepackage{amsfonts}
\usepackage{graphicx}
\usepackage{epstopdf}
\usepackage{algorithmic}
\usepackage{dsfont}
\ifpdf
  \DeclareGraphicsExtensions{.eps,.pdf,.png,.jpg}
\else
  \DeclareGraphicsExtensions{.eps}
\fi

\usepackage{enumitem}
\setlist[enumerate]{leftmargin=.5in}
\setlist[itemize]{leftmargin=.5in}

\title{Optimal Trading in Automated Market Makers with Deep Learning
\thanks{
S.J. acknowledges the support of the Natural Sciences \& Engineering Research council of Canada through NSERC Alliance [ALLRP 550308 - 20] and the University of Toronto's Data Sciences Institute.}}

\author{ Sebastian Jaimungal\thanks{University of Toronto and Oxford-Man Institute, Oxford University (\email{sebastian.jaimungal@utoronto.ca},\protect\url{http://sebastian.statistics.utoronto.ca})}
\and Yuri Saporito\thanks{Escola de Matem\'atica Aplicada, Funda\c{c}\~ao Getulio Vargas, Brazil (\email{yuri.saporito@fgv.br},\protect\url{http://yurisaporito.com})}
\and Max de Souza\thanks{Instituto de Matem\'atica e Estat\'istica, Universidade Federal Fluminense, Brazil (\email{maxsouza@id.uff.br})}
\and Yuri Thamsten\thanks{Instituto de Matem\'atica e Estat\'istica, Universidade Federal Fluminense, Brazil (\email{ythamsten@id.uff.br})}
}

\usepackage{amsopn}

\DeclareMathOperator*{\argmin}{argmin}
\newcommand{\E}{\mathbb{E}}

\newcommand{\vx}{{\boldsymbol x}}

\newcommand{\vb}{{\boldsymbol b}}

\newcommand{\vu}{{\boldsymbol u}}

\newcommand{\vw}{{\boldsymbol w}}

\newcommand{\vy}{{\boldsymbol y}}

\newcommand{\A}{{\mathcal{A}}}
\newcommand{\tT}{{t\in[0,T]}}

\makeatletter
\newcommand*{\addFileDependency}[1]{
  \typeout{(#1)}
  \@addtofilelist{#1}
  \IfFileExists{#1}{}{\typeout{No file #1.}}
}
\makeatother

\graphicspath{{figures/empirical/}{figures/dgm}{figures/ddqn}}

\usepackage{setspace}
\usepackage{todonotes}

\begin{document}

\maketitle

\begin{abstract}
This article explores the optimisation of trading strategies in Constant Function Market Makers (CFMMs) and centralised exchanges. We develop a model that accounts for the interaction between these two markets, estimating the conditional dependence between variables using the concept of conditional elicitability. Furthermore, we pose an optimal execution problem where the agent hides their orders by controlling the rate at which they trade. We do so without approximating the market dynamics. The resulting dynamic programming equation is not analytically tractable, therefore, we employ the deep Galerkin method to solve it. Finally, we conduct numerical experiments and illustrate that the optimal strategy is not prone to price slippage and outperforms na\"ive strategies.
\end{abstract}


\section{Introduction}


Automated Market Makers (AMMs) and, more specifically, Constant Function Market Makers (CFMMs), including Uniswap \cite{adams2020uniswap}, \cite{adams2021uniswap},  now dominate the  decentralised finance (DeFi) landscape. In contrast to centralised exchanges, where liquidity providers (LPs) post volumes and rates that they are willing to buy/sell at that rest in a limit order book (LOB) until matched by liquidity takers (LTs), CFMMs allow tokens to be traded through a prespecified formula. In particular, CFMMs transact such that a \textit{bonding function} or \textit{invariant} $f$ remains constant before and after the trade. The invariant $f$ maps the reserves $(r^x,r^y)$ of token-$x$ and token-$y$, respectively, to a real number, e.g., in a constant product market maker (CPMM) the invariant is $f(r^x,r^y) = r^x\,r^y$. 

Several recent publications have contributed to the study of optimal trading in CFMMs. In \cite{cartea2022decentralised-es}, the authors present models for LTs that operate in a CPMM when prices are formed in the pool and in alternative venues. In \cite{cartea2022decentralisedl-llp}, the authors examine liquidity provision in CFMMs and in CPMMs with concentrated liquidity (i.e., where traders provide liquidity when the marginal price is within set bounds) and derive optimal LP strategies under some approximations. 
The work of \cite{bergault2022automated} proposes a new design for AMMs using traditional market making models. 
The optimal allocation of liquidity for LPs who hold beliefs on the future price movements is analysed in \cite{neuder2021strategic} and \cite{fan2022differential} -- with implications for the design of smart contracts. \cite{angeris2022constant}  and \cite{angeris2022optimal} use convex optimisation to explore routing and multi-asset trading in CFMMs. Lastly, and most closely related to this work, \cite{cartea2023execution} explore statistical arbitrage strategies when trading in AMMs. The main differences with our work and the latter is that while \cite{cartea2023execution} employ various approximations in the dynamics and resulting dynamic programming equations to render the solutions analytically tractable,
 here we deploy deep learning tools to determine the optimal strategies.

The problems we address here are two fold. Firstly, we develop a model for how trades in a centralised exchange and those in a related CFMM interact with one another. This contrasts with much of the extant literature that assume the markets interact in a single direction. Moreover, we model the individual trades and do not approximate them as continuous processes. We use the notion of conditional elicitability (see \cite{fissler2016higher} and \cite{coache2022conditionally} for its use in reinforcement learning) to estimate the conditional dependence between variables so as not to enforce a pre-determined structure on it. 
Secondly, we pose and solve an optimal execution problem, where the trader has an initial amount of token-$x$ that they wish to liquidate into token-$y$. They can do so by sending trades to a CFMM.
We pose this as a stochastic control problem, without making approximations on the dynamics, and then solve the resulting Hamilton-Jacobi-Bellman (HJB) using the deep Galerkin method (DGM) \cite{sirignano2018dgm}, see also \cite{al2022extensions} and  \cite{al2018solving}. We utilise DGM, as the resulting HJB equation is not analytically tractable which stems from the complex market dynamics that we do not approximate. Furthermore, the problem dimension is more than three, which makes standard numerical PDE solvers not applicable.

\section{Empirical Analysis}

Our empirical study centres around the ETH-USDC pool of \allowbreak Uniswap v3 in the month of September, 2022. This particular pool is the one that invariably has the highest trading volume. To serve as a reference spot counterpart, we utilise Level 1 data from Binance's ETH-BUSD order book over the same time frame. We choose this order book as it typically exhibits high daily traded volume compared to others. 
\begin{wrapfigure}{r}{0.7\textwidth}%
    \centering
    \includegraphics[width=0.7\textwidth]{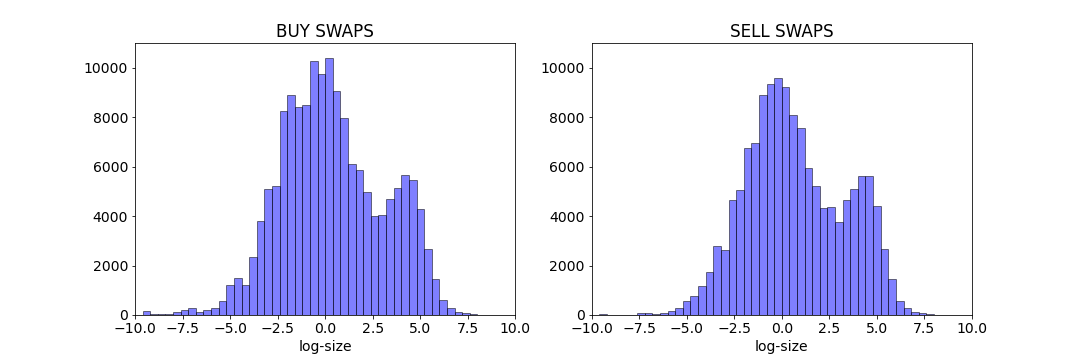}%
    \caption{Histograms of the log-sizes of swaps occurring in the pool.}%
    \label{fig:sizes}%
\end{wrapfigure}
Additionally, as both \allowbreak BUSD and USDC are USD-stablecoins, the fact that ETH is traded \allowbreak against USDC in the pool and BUSD in the exchange does not alter our analysis. In fact, USDC deposits on Binance are automatically converted to BUSD in a 1-1 fashion, and BUSD (or any other USD stablecoin, with the exception of USDT) can be withdrawn as USDC within the same basis\footnote{See, e.g., \url{https://tinyurl.com/bdhx4ybn}}. We denote the pool price (in the CPMM) and the spot price (in the centralised exchange) as $P_t$ and $S_t$, respectively. 




\subsection{Swaps sizes}

We start by creating histograms of the log-size of swaps, as shown in Figure \ref{fig:sizes}. To simplify the discussion, we establish ETH as the base currency and USDC as the quote currency for the pool. This allows us to categorise swaps as either ``buy'' or ``sell''. Specifically, a ``buy'' swap involves exchanging USDC for ETH, while a ``sell'' swap involves trading ETH for USDC. The figures demonstrate that the volume of swaps do not follow a log-normal distribution, and instead displays a bimodal structure. This suggests the possibility of two distinct trading regimes.

\subsection{Cross-exciting nature of swap flow}

\begin{wraptable}{r}{0.5\textwidth}
\footnotesize
  \centering
  \vspace{-1em}
    \begin{tabular}{rrrrr}
    \toprule\toprule
          & \multicolumn{1}{l}{$\tau_{buy|buy}$} & \multicolumn{1}{l}{$\tau_{sell|buy}$} & \multicolumn{1}{l}{$\tau_{buy|sell}$} & \multicolumn{1}{l}{$\tau_{sell|sell}$} 
          \\[0.25em]
    \midrule
    mean  & 16.23 & 14.29 & 11.76 & 18.38 
    \\
    std.dev.   & 21.77 & 25.54 & 21.66 & 24.43 
    \\
    25\% & 0     & 0     & 0     & 0 
    \\
    50\% & 12.00    & 0     & 0     & 12.00 \\
    75\% & 24.00    & 24.00    & 13.00    & 24.00 
    \\
    $\lambda$ & 0.037 & 0.070 & 0.085 & 0.034
    \\
    \bottomrule
    \bottomrule
    \end{tabular}%
    \vspace*{0.25em}
    \caption{Empirical distribution of inter-arrival times between swap events conditional on the type of swap event.}
    \vspace*{-1.5em}
  \label{tab:tau-swap}%
\end{wraptable}

Next, we examine the cross-excitation in swap flow. We present Table \ref{tab:tau-swap} to support this hypothesis. The table reports the mean, standard deviation, and 25\%, 50\%, and 75\% quantiles of the inter-arrival times between consecutive buy events, a sell followed by a buy, a buy followed by a sell, and consecutive sell events. 

As reported, buy swaps tend to arrive more rapidly after sell swaps than after buy swaps suggesting the cross-exciting nature of swap arrivals. This may be due to the potential for an arbitrage opportunity created by the arrival of a swap of a certain size that upsets the pool's reserves.

\begin{figure}[htb!]%
    \centering
    \includegraphics[width=0.79\textwidth]{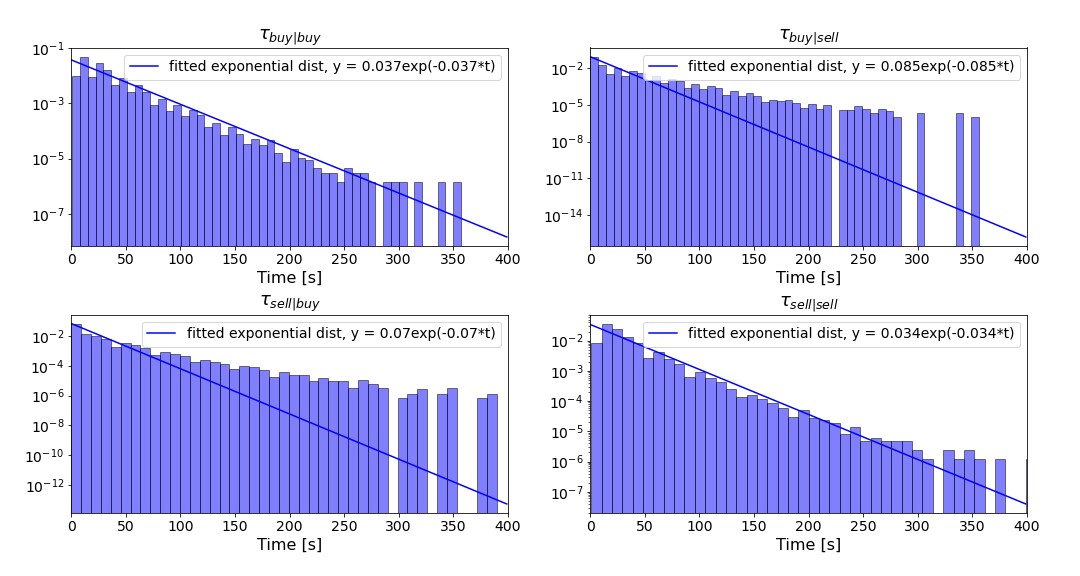} %
    \caption{Inter-arrival times between buy and sell swap events.}%
    \label{fig:swaps-interarrival-times}%
\end{figure}

Figure \ref{fig:swaps-interarrival-times} displays histograms of the inter-arrival times of swaps, along with an exponential fit to the inter-arrival times. The inter-arrival times for buy-buy and sell-sell swaps are approximately exponentially distributed, while the inter-arrival times for buy-sell and sell-buy swaps exhibit heavier tails, potentially due to the cross-excitation previously discussed. The estimated arrival rates $\lambda$ are provided in Table \ref{tab:tau-swap}, which were estimated by curve fitting of the histograms shown in Figure \ref{fig:swaps-interarrival-times}.
\FloatBarrier





\subsection{Mean-reversion of Spot-Pool spread}

Our focus now shifts to examining price changes in both the AMM and the centralised exchange. To accomplish this, we introduce the concept of pool and centralised arithmetic returns:
$$
r^P_t := P_t - P_{t^-} \text{ and } r^S_t := S_t - S_{t^-},
$$
where, as usual, $a_{t^-}=\lim_{s\uparrow t} a_s$ for any process $a$.
At each swap arrival time $t$, we record the pool price $P_t$, which equals  the exchange rate of the corresponding swap. As for $S_t$, we sample the spot price at each trade time and record the exchange's mid-quote immediately before that point in time.  When conditioning trades to the spot-pool spread, we reduce the sample to times when swap prices change. This is necessary as many trades in the centralised exchange tend to occur within a one-second window, whereas the time scale for swaps in the AMM is seconds.

\begin{figure}[!htbp]%
    \centering
    \includegraphics[width=0.7\textwidth]{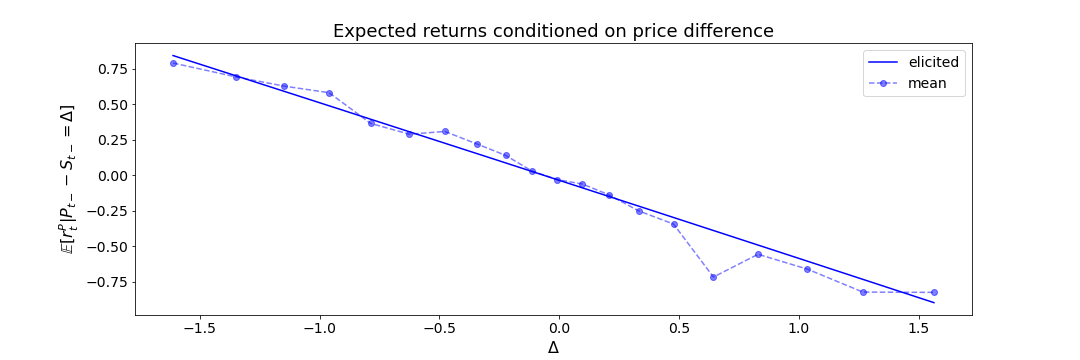}%
    \qquad
    \includegraphics[width=0.7\textwidth]{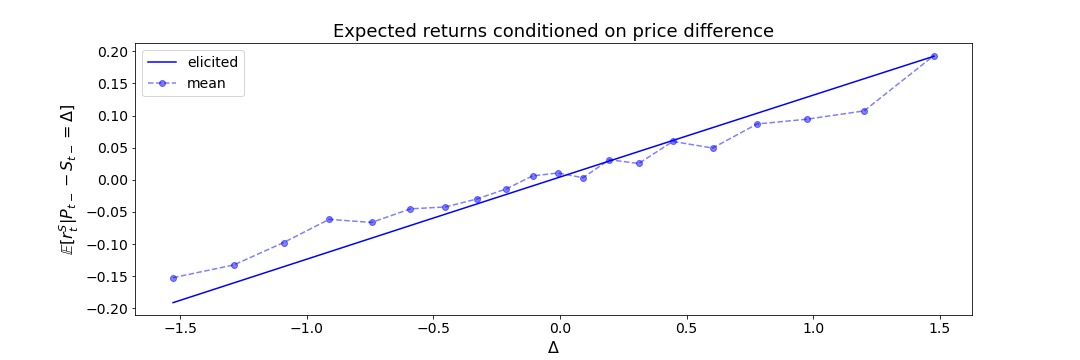} %
    \caption{Conditional expected return in each venue conditioned on a certain price differential value.}%
    \label{fig:expec-rets-cond-diff}%
\end{figure}

In Figure \ref{fig:expec-rets-cond-diff}, we present an estimate of the expected returns conditioned on a spread, $\E[r^P|P-S]$ and $\E[r^S|P-S]$, using two methods. First, we bin the data into those with spread $\Delta = P-S \in\{(-1.75,-1.5], (-1.5,-1.25], \dots, (1.5,1.75]\}$, and compute the mean return $r^P$ and $r^S$ within each of these bins. This provides us with the dash-dot line in the figure. Second, we estimate the conditional mean by projecting the data onto a Legendre polynomial basis up to order five. This provides us with the solid line in the figure. As expected, the projection approach provides a smoother representation of the estimate of the conditional expectation. Moreover, the estimate suggests there is a linear relationship between returns and the spread, with pool returns decreasing as the spread decreases, and spot returns increasing as the spread increases.

To evaluate the reliability of our estimates, we employ the projection method on randomly selected subsets of our dataset. Each trial consists of a subset of size equal to 80\% of the full dataset, and we repeat this experiment 300 times. The resulting estimates  are presented in the appendix in Figure \ref{fig:expec-rets-cond-diff-many-regs} and illustrates that the estimated dependence is stable across sub-samples. 


\subsection{Arbitrage-driven spot-pool order flow}\label{sec:elicitability}

\begin{figure}[t]%
    \centering
    \begin{minipage}[t]{0.45\textwidth}
    \centering
    \begin{tikzpicture}
        \draw (0, 0) node[inner sep=0] {\includegraphics[width=\textwidth]{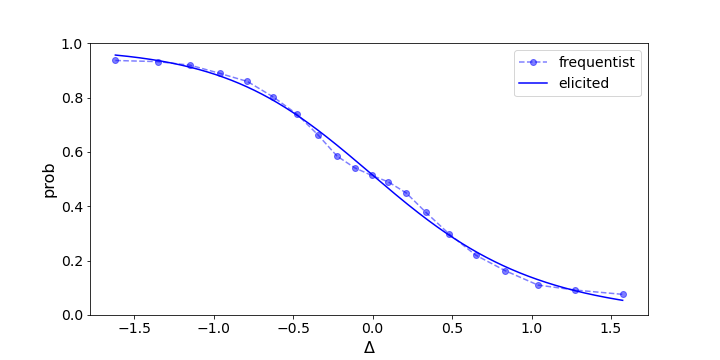}};
        \draw (0, 1.6) node {
        \footnotesize$\mathbb{P}(\text{buy}\,|\,\text{AMM swap}, \,\Delta=P-S)$};
    \end{tikzpicture}
    \vspace*{-2em}
    \caption{Conditional probability that an AMM swap is a buy conditional on the spread.}%
    \label{fig:swaps-probs-cond-diff}%
    \end{minipage}
    \quad
    \begin{minipage}[t]{0.45\textwidth}
    \centering
    \begin{tikzpicture}
        \draw (0, 0) node[inner sep=0] {\includegraphics[width=\textwidth]{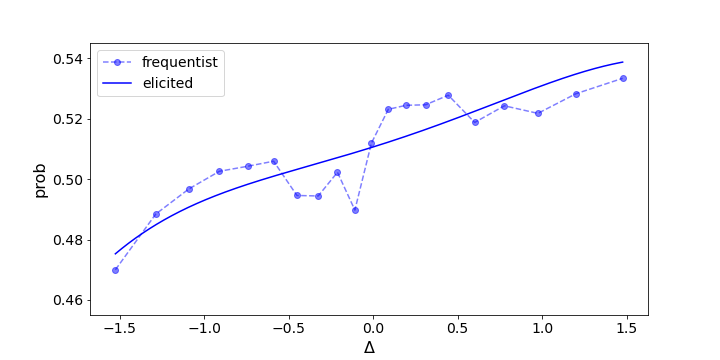}};
        \draw (0, 1.6) node {\footnotesize$\mathbb{P}(\text{buy}\,|\,\text{Ref trade}, \,\Delta=P-S)$};
    \end{tikzpicture}
    \vspace*{-2em}
    \caption{Conditional probability that a trade in the centralised exchange is a buy conditional on the spread.}%
    \label{fig:trade-probs-cond-diff}%
    \end{minipage}
\end{figure}
Next, we focus on modelling the probability of a swap being a buy or sell conditional on the spread. To this end, Figure \ref{fig:swaps-probs-cond-diff} shows two estimates of this conditional probability. The first is again by binning the data using the same spread bins as in the previous section, namely, $\Delta = P-S \in\{(-1.75,-1.5], (-1.5,-1.25], \dots, \allowbreak(1.5,1.75]\}$, and computing the empirical probability of a buy/sell swap occurring. This provides us with the dash-dot lines in the figures. The second approach uses elicitable maps \cite{fissler2016higher,coache2022conditionally}. Specifically, if we let $\mathcal{G}$ denote the class of all functions $\mathds{R}\to[0,1]$, then define
\begin{equation}
    g^*=\argmin_{g\in\mathcal{G}} \E\left[ 
    \Big(g(\Delta_t) - \mathds1_{swap_t=buy} \Big)^2
    \right]\,. \label{eqn:elicitable-map}
\end{equation}

This minima is unique, and the minimiser $g^*(x)$ then equals $\mathbb{P}(swap_t=buy|\Delta_t=x)$, and the ``score'' function is said to elicit the conditional probability. Similarly, when the event $swap_t=buy$ is replaced by $swap_t=sell$. When the expectation is replaced with the empirical mean, then the optimiser of \eqref{eqn:elicitable-map} is an estimator of the conditional probability. In practice, it is not possible to seek over the full set $\mathcal{G}$, rather, we seek over functions $g=(1+e^{-h})^{-1}$, where $h\in\mathcal{L}$, and $\mathcal{L}$ is the set of functions spanned by Legendre polynomials up to order four. This results in the solid lines in the figures.

These results can be interpreted in terms of arbitrage flows. When the spread $\Delta=P-S$ is significantly positive (or negative), i.e., of sufficient magnitude in absolute value, arbitrageurs will engage in selling (or buying) in the pool and buying (or selling) in the spot market. The estimates we obtain and show in Figure \ref{fig:swaps-probs-cond-diff} reflect this effect precisely.

To evaluate the reliability of our estimates, we employ the projection method on randomly selected subsets of our dataset. As before, each trial consists of a subset of size equal to 80\% of the full dataset, and we repeat this experiment 300 times. Figure \ref{fig:swap-probs-cond-diff-uncertainty} shows the difference of these sub-sample estimates and the estimate from the full data set. For most of the estimates, the deviation is less than $0.002$ in magnitude, indicating that the reliability of the estimated conditional expectations in Figure \ref{fig:swaps-probs-cond-diff}.

Figure \ref{fig:trade-probs-cond-diff} shows the analogous estimates for the centralised exchange. In particular, it shows the probability that when a trade arrives in the centralised exchange that the trade is a buy, conditional on the spread.
Once again, we evaluate the reliability of the estimated dependence by repeating the exercise on 300 sub-samples of 80\% of the data. The difference between the estimated dependence and the one shown  in Figure \ref{fig:trade-probs-cond-diff} are reported in Figure \ref{fig:trade-probs-cond-diff-uncertainty}. Once again, the errors are mostly less than $0.002$ in magnitude, indicating that the estimated dependence is accurate.

\begin{wraptable}{r}{0.5\textwidth}
\footnotesize
  \centering
  \vspace{-1em}
  \caption{Coefficients of Legendre basis functions for estimated conditional probability of buy events}
    \begin{tabular}{ccc}
    \toprule
    \toprule
    \multicolumn{1}{c}{$i$} & \multicolumn{1}{l}{AMM Swap  } & \multicolumn{1}{l}{Centralised Trade} \\
    & $a_i$ & $b_i$
    \\
    \midrule
    0     & 0.1154 & 0.0190 \\
    1     & -3.3510 & 0.2065 \\
    2     & 0.0010 & -0.0944 \\
    3     & 0.1123 & 0.0268 \\
    4     & -0.0717 & -0.0586 \\
    \midrule
    shift $\theta$ & 0.0153 & -0.0377 \\
    scale $h$ & 1.7635 & 2.5800 \\ 
    arrival rate $A$ & 0.4166 & 0.0833 \\
    \bottomrule
    \bottomrule
    \end{tabular}%
    \vspace*{-5em}
  \label{tab:swap-trade-Legendre-coeffs}%
\end{wraptable}%
The estimated coefficients for both swaps and trades are reported in Table \ref{tab:swap-trade-Legendre-coeffs}. For both event types, the spread is first shifted and scaled (as reported in the table) before passing to the Legendre polynomial basis.

\section{Continuous-Time Formulation}

As anticipated, the preceding section demonstrates significant inter-dependencies between the AMM and the centralised exchange. With these findings, we now have a reliable quantitative foundation upon which to construct a continuous-time model that could potentially be employed to optimise trading strategies.

Let the base crypto-currency be denoted $x$, the alternate crypto-currency by $y$ and $(r_t^{x,y})_{t\ge0}$ denote the reserve processes of $x,y$ in the AMM pool, respectively. The resulting marginal price is therefore $P_t = r^y_t/r^x_t$. Further, let the centralised exchange mid-price process of $y$ in terms of $x$ be denoted $(S_t)_{t\ge0}$. Building on the empirical discussion above, we propose the market model described in the next sections.

\subsection{Reference Spot price Model}  
The counting processes and associated intensity processes for trades inducing an upward (+) / downward (-) tick in the reference market are denoted by $(M_t^\pm){t\ge0}$ and $(\kappa_t^\pm){t\ge0}$, respectively. Based on the empirical evidence in Section \ref{sec:elicitability}, we model the intensity processes as
\begin{subequations}
    \begin{align}
        \kappa_t^+ = \kappa^+\left(\frac{r^y_t}{r_t^x} - S_t\right)
        &:= A_\kappa\,
        \Big(1+e^{- f^+_\kappa(r_t^y/r_t^x - S_t)} \Big)^{-1}\,,
        \qquad \text{and}
        \\
        \kappa_t^- = \kappa^-\left(\frac{r^y_t}{r_t^x} - S_t\right)
        &:= A_\kappa - \kappa^+\left(\frac{r^y_t}{r_t^x} - S_t\right)
    \end{align}%
\end{subequations}
where $f_\kappa^+(\Delta) = \sum_{i=0}^{4} b_i \;L_i\left(\frac{\Delta+\theta_b}{h_b}\right)$, $L_i$ is the Legendre polynomial of order $i$ and $A_\kappa$ is the baseline arrival rate of trades in the reference market. The various estimates for the coefficients are as reported in Table \ref{tab:swap-trade-Legendre-coeffs}.
    
\subsection{AMM Swap Arrival Model} 
The counting processes and intensities for swap events in the AMM pool of swapping $x$ for $y$ and $y$ for $x$ are denoted by $(N_t^{x,y}){t\ge0}$ and $(\lambda_t^{x,y}){t\ge0}$, respectively. As before, based on the empirical evidence presented in Section \ref{sec:elicitability}, we model the intensity processes as
     \begin{subequations}
        \begin{align}
            \lambda_t^x = \lambda^x\left(\frac{r^y_t}{r_t^x} - S_t\right)
            &:= A_\lambda\;\Big(1+e^{- f^x_\lambda(r_t^y/r_t^x - S_t)}\Big)^{-1}\,,  
            \qquad \text{and}
            \\
            \lambda_t^y = \lambda^y\left(\frac{r^y_t}{r_t^x} - S_t\right)
            &:= A_\lambda - \lambda^x\left(\frac{r^y_t}{r_t^x} - S_t\right)
        \end{align}%
    \end{subequations}%
where $f_\lambda^x(\Delta) = \sum_{i=0}^{4} a_i \;L_i\left(\frac{\Delta+\theta_a}{h_a}\right)$  and $A_\lambda$ is the baseline arrival rate of swaps in the pool. The various estimates for the coefficients are as reported in Table \ref{tab:swap-trade-Legendre-coeffs}.

\subsection{Joint Dynamics of AMM and Reference Market}
When a swap event occurs, we must ensure the invariant remains so before and after a swap. Hence, if we swap $\pi^y$ of token-$y$ for token-$x$, accounting for fees at a rate of $\phi$, we receive $\frac{r^x(1-\phi)\pi^y}{r^y+(1-\phi)\pi^y}$ of token-$x$. Similarly, if we swap $\pi^x$ of token-$x$ for token-$y$,  we receive $\frac{r^y(1-\phi)\pi^x}{r^x+(1-\phi)\pi^x}$ of token-$y$. We can account for both swap transactions by defining the swap function 
\begin{equation}
    f(a,b,\pi) = \frac{b\,\pi\,(1-\phi)}{a + \pi (1-\phi)}.
\end{equation}    
In this case, we receive $f(r^y,r^x,\pi^y)$ of token-$x$ when we swap $\pi^y$ of token-$y$ and we receive $f(r^x,r^y,\pi^x)$ of token-$y$ when we swap $\pi^x$ of token-$x$.

\begin{figure}[H]
    \centering
    \includegraphics[width=0.8\textwidth]{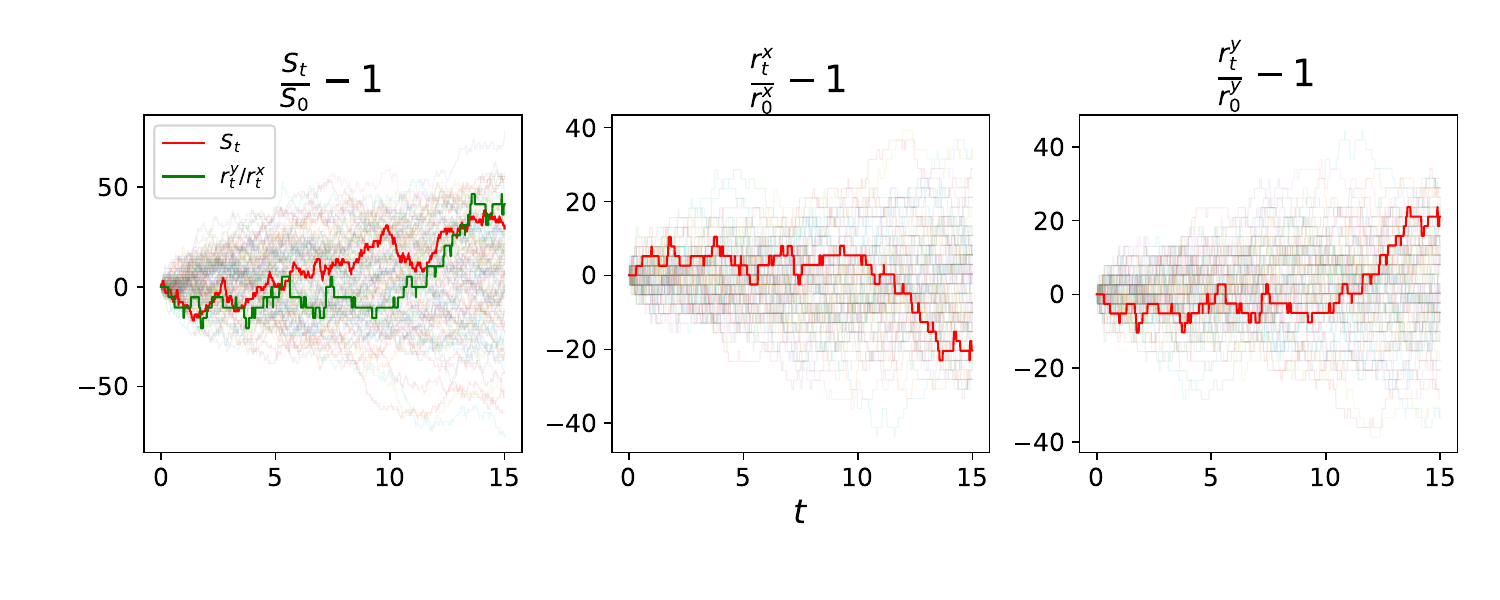}
    \vspace*{-3em}
    \caption{Sample paths of cumulative returns of the uncontrolled market dynamics over 15 minutes $(\times 10^{-5})$.}
    \label{fig:sample-paths}
\end{figure}

To construct a joint model, we assume that swap events are always for a fixed amount of token-$x$ or token-$y$, denoted by $\pi^x>0$ and $\pi^y>0$, respectively. Additionally, we assume that the centralised exchange price moves in fixed ticks of $\theta^\pm$ up or down. While it is possible to generalise this assumption to cases where trade volumes and tick movements are random, we leave this to future work. Having modelled the interaction between the centralised exchange movements and AMM swap events, we are now able to present the joint dynamical model explicitly. This includes the uncontrolled market dynamics for the spot price $S$, token-$x$ reserves $r^x$, and token-$y$ reserves $r^y$, which are governed by the following stochastic differential equations (SDEs):
\begin{align}\label{eq:model_uncontrolled}
\begin{cases}
    dS_t &= \theta^+\,dM_t^+ - \theta^-\,dM_t^-\,,
    \\[0.5em]
    dr_t^x &= \pi^x\,dN_t^x - f(r^y_t, r^x_t,\pi^y)\,dN_t^y\,,
    \\[0.5em]
    dr_t^y &= - f(r^x_t, r^y_t,\pi^x)\,dN_t^x + \pi^y\,dN_t^y\,. 
\end{cases}
\end{align}
As previously modelled, the intensity processes for the various counting processes exhibit complex interdependent relationships and mutual excitation. Moreover, the occurrence of events themselves can alter the state of the system, subsequently modifying the arrival rates of events and propagating the intricate interplay between the intensity processes. Figure \ref{fig:sample-paths} shows several sample paths together with a single highlighted path and illustrates the interacting nature of the processes. In the left panel,  $\frac{r_t^y}{r_t^x}$ is the AMM's marginal price. 

\subsection{Agent's Optimisation Problem} Next, we pose the optimisation problem that the agent wishes to solve. For this purpose, let $(L_t)_\tT$ denote the counting process corresponding to the agent's swap orders of $x$ for $y$ (all assumed of size $\zeta$) over the trading period $[0,T]$. The agent controls the intensity  $(\ell_t)_\tT$ of the counting process, but does not control exactly when the trade is sent to the AMM. The reason we model trades this way is so that the agent's trade times are randomised, and hence the trader is able to partially hide their intentions. As trade times are randomised, we add a single trade at the end of the trading horizon to ensure full liquidation. 

The market dynamics \eqref{eq:model_uncontrolled} must be modified to account for the trading actions of the agent in the AMM pool. Accounting for the agent's trading actions results in the system
\begin{align}
\begin{cases}
dS_t &= \theta^+\,dM_t^+ - \theta^-\,dM_t^-\,,
\\[0.5em]
dr_t^x &= \pi^x\,dN_t^x - f(r^y_t, r^x_t,\pi^y)\,dN_t^y + \zeta\,dL_t\,,
\\[0.5em]
dr_t^y &= - f(r^x_t, r^y_t,\pi^x)\,dN_t^x + \pi^y\,dN_t^y - f(r^x_t,r_t^y,\zeta)\,dL_t\,.
\end{cases}
\label{eqn:market-sde-with-trader}
\end{align}
The last term in the second and third lines correspond to the modification due to the agent's trading actions. We assume the trader uses only the AMM to complete the trade.
The admissible trading strategies can be modified to allow for swapping token-$y$ for token-$x$, in addition to swapping token-$x$ for token-$y$. In this case, the problem becomes more related to that of an agent engaging in statistical arbitrage. This modification is left for future work.

We wish, instead, to solve a problem where the agent has initial inventory of $Q$ in crypto-$x$ and $0$ in crypto-$y$. We denote the agent's inventory in crypto-$x$ and crypto-$y$ at time $t$ by $(z^{x,y}_t)_\tT$, respectively. The dynamics of these inventories are given by
    \begin{align}
        dz_t^x &= -\zeta\,dL_t,
        \\[0.5em]
        dz_t^y &= f(r^x_t, r^y_t,\zeta)\,dL_t,\label{eq:zy}
    \end{align}
with $z_0^x = Q$ and $z_0^y = 0$. The agent wishes to liquidate crypto-$x$ and obtain crypto-$y$ and maximise the total amount of crypto-$y$ obtained during the liquidation. Specifically, they wish to compute
\begin{equation}
    \sup_{\ell\in\A}\E\left[\beta\left( z_T^y + f(r_T^x,r_T^y,\, z_T^x)\right) - \alpha\, (z_T^x)^2 - \frac{\varphi}{2} \int_0^T \ell_t^2 dt \right]\,,
    \label{eqn:optimisation-problem}
\end{equation}
where $\A$ is the space of admissible intensity processes, meaning positive, progressively measurable processes such that $\E[\int_0^T \ell_t^2\,dt]<\infty$. 

The first term represents the agent's token-$y$ inventory by the end of the trading horizon, the second term represents the lump-sum of token-$y$ they receive by liquidating all remain token-$x$ at $T$, the third term represents a penalty on having remaining shares to liquidate, and the fourth term penalises trading to quickly. 


Due the complexities in the market dynamics \eqref{eqn:market-sde-with-trader}, the optimisation problem \eqref{eqn:optimisation-problem} cannot be solved analytically. Instead, in the next section, we employ deep learning techniques to approximate the optimal strategy.

\section{Numerical Implementation}

In this section, we utilise the Deep Galerkin Method (DGM)  (\cite{sirignano2018dgm}, \cite{al2022extensions}, and \cite{al2018solving}) to approximate the solution to the optimisation problem \eqref{eqn:optimisation-problem}. To accomplish this, we derive the dynamic programming equations associated with the optimisation problem, resulting in a non-linear partial-integro differential equation (PIDE). We then apply DGM to solve the PIDE, and obtain the optimal trading strategy in feedback form.

To do so, first we introduce the value function associated with
\eqref{eqn:optimisation-problem},
\begin{align*}
\small
    &V(t,S,r^x,r^y,z^x,z^y) \\
    &= \sup_{\ell\in\A[t,T]}\E_t\left[\beta\left( (z_T^y)^{t,z^y} 
    + f\big((r_T^x)^{t,r^x},(r_T^y)^{t,r^y},\, (z_T^x)^{t,z^x}\big) \right)- \alpha\, ((z_T^x)^{t,z^x})^2 - \frac{\varphi}{2} \int_t^T \ell_s^2ds \right],
\end{align*}
where the notation $a_s^{t,a}$ denotes the value of the process $(a_u)_{u\ge t}$ at time $s$ when it starts at the value $a$ at time $t$, and $\A[t,T]$ is the space of admissible intensities on the interval $[t,T]$. Under well-known regularity conditions (see e.g., \cite{oksendal2007applied}), the value function solves the non-linear Hamilton-Jacobi-Bellman (HJB) equation\footnote{To simplify the presentation, we elect to suppress the arguments of $\partial_t V$ and $\Delta_\cdot V$.}
\begin{equation}
\begin{split}
\label{eqn:HJB-1}
\partial_t V &+ \kappa^+\!\big(\tfrac{r^y}{r^x} - S\big)\;\Delta_S^+V
+  \kappa^-\!\big(\tfrac{r^y}{r^x} - S\big)\;\Delta_S^- V 
\\
&+ \lambda^x\!\big(\tfrac{r^y}{r^x} - S\big)\;\Delta_x V +  \lambda^y\!\big(\tfrac{r^y}{r^x} - S\big)\;\Delta_y V 
+\sup_{\ell \geq 0} \left\{  \Delta_\zeta V \,\ell - \frac{\varphi}{2} \,\ell^2 \right\}= 0,
\end{split}
\end{equation}
with final condition $V(T,S,r^x,r^y,z^x,z^y) = \beta\left( z^y + f(r^x,r^y,z^x) \right) - \alpha (z^x)^2$ and where the difference operators are defined as
\begin{align*}
    \allowdisplaybreaks
    \Delta_S^+V(t,S,r^x,r^y,z^x,z^y)&:=V\big(t,S+\theta^+, r^x,r^y,z^x,z^y\big)  - V\big(t,S, r^x,r^y,z^x,z^y\big)
    \\[0.25em]
    \Delta_S^-V(t,S,r^x,r^y,z^x,z^y)&:=V\big(t,S-\theta^-, r^x,r^y,z^x,z^y\big)- V\big(t,S, r^x,r^y,z^x,z^y\big)
    \\[0.25em]
    \Delta_xV(t,S,r^x,r^y,z^x,z^y)&:=V\big(t,S, r^x+\pi^x,r^y-f(r^x,r^y,\pi^x),z^x,z^y\big)- V\big(t,S, r^x,r^y,z^x,z^y\big)
    \\[0.25em]
    \Delta_yV(t,S,r^x,r^y,z^x,z^y)&:=V\big(t,S,r^x - f(r^y,r^x,\pi^y), r^y + \pi^y,z^x,z^y\big)- V\big(t,S, r^x,r^y,z^x,z^y\big)
    \\[0.25em]
    \Delta_\zeta V(t, S, r^x, r^y, z^x, z^y)&:=V\left(t,S,r^x + \zeta, r^y - f(r^x,r^y,\zeta), z^x - \zeta, z^y + f(r^x,r^y,\zeta)\right)\\
    &\phantom{:=}- V\big(t,S, r^x,r^y,z^x,z^y\big).
\end{align*}
The feedback form of the optimal control, obtained by finding the $\ell$ that attains the supremum in \eqref{eqn:HJB-1}, is 
\begin{align*}
\ell^*(t, S, r^x, r^y, z^x, z^y) = 
\frac{1}{\varphi}\Big(\Delta_\zeta V(t, S, r^x, r^y, z^x, z^y)\Big)_+,
\end{align*}
where $(x)_+ := \max\{x,0\}$. Upon substituting this feedback form into the HJB equation, we obtain the PIDE
\begin{align*}
\partial_t V &+ \kappa^+\!\big(\tfrac{r^y}{r^x} - S\big)\;\Delta_S^+V 
+  \kappa^-\!\big(\tfrac{r^y}{r^x} - S\big)\;\Delta_S^- V 
\\
&+ \lambda^x\!\big(\tfrac{r^y}{r^x} - S\big)\;\Delta_x V  +  \lambda^y\!\big(\tfrac{r^y}{r^x} - S\big)\;\Delta_y V 
+\frac{1}{2\varphi} \big(\Delta_\zeta V\big)_+^2 = 0.
\end{align*}
The last term, which is non-linear, and the terminal condition prevents us from being able to solve the PIDE in closed-form. We can, however, reduce the dimension of the problem by one. 

\paragraph{Dependence on $z^y$} We make use of the structure of the model to show that $V$ depends linearly on $z^y$. Due to the dynamics of $z^y$ in \eqref{eq:zy} and that the terminal condition in linear in $z^y$, we find that 
\begin{align}
\begin{split}
        V(t,S,r^x,r^y,z^x,z^y) = 
    \beta\,z^y &+\sup_{\ell\in\A[t,T]}\E_t\bigg[\beta\,
    f\Big((r_T^x)^{t,r^x},(r_T^y)^{t,r^y},\, (z_T^x)^{t,z^x}\Big) 
    - \alpha \Big((z_T^x)^{t,z^x}\Big)^2
    \\
    &\hspace*{6em} 
    +  \int_t^T \beta\,f\Big((r_s^x)^{t,r^x}, (r_s^y)^{t,r^y}, \zeta\Big)\,dL_s 
    - \frac{\varphi}{2} \int_t^T \ell_s^2ds 
    \bigg]\,.
\end{split}
\label{eqn:zy-to-running-cost}
\end{align}
Therefore, we may write
$$V(t,S,r^x,r^y,z^x,z^y) = \beta\,z^y + v(t,S,r^x,r^y,z^x),$$
where $v$ satisfies the PIDE
\begin{align}
\begin{split}
\partial_t v &+ \kappa^+\!\big(\tfrac{r^y}{r^x} - S\big)\;\Delta_S^+v
+  \kappa^-\!\big(\tfrac{r^y}{r^x} - S\big)\;\Delta_S^- v 
\\
& 
+ \lambda^x\!\big(\tfrac{r^y}{r^x} - S\big)\;\Delta_x v  
+  \lambda^y\!\big(\tfrac{r^y}{r^x} - S\big)\;\Delta_y v 
+\tfrac{1}{2\varphi} \big(\beta\,f(r^x,r^y,\zeta) + \Delta_\zeta v\big)_+^2 = 0.
\end{split}
\label{eqn:PIDE-v}
\end{align}
with terminal condition 
\begin{align}\label{eq:terminal}
v(T,S,r^x,r^y,z^x) =  \beta\,f(r^x,r^y,z^x) - \alpha \,(z^x)^2,
\end{align}
and
\begin{align}\label{eq:optimal_control}
\ell^*(t, S, r^x, r^y, z^x) = \frac{1}{\varphi}
\Big(\beta\,f(r^x,r^y,\zeta) + (\Delta_\zeta v ) \Big)_+.
\end{align}
The $\Delta$ difference operators are similar to the ones previously defined. In particular, the $\Delta_\zeta$ obviously does not consider the difference with respect to $z^y$.

\paragraph{Implementation details} To ensure the DGM algorithm can efficiently solve the PIDE \eqref{eqn:PIDE-v}, we normalise the variables such that, during the DGM training, the normalised variables can be sampled within $[0,1]$. Specifically, we rescale the various entries of $v$ and work with the function
\[
\tilde{v}(s,x,a,b,c) := v(T\,s, \bar{S}\,x, \bar{r}\,a, \bar{S} \,\bar{r}\,b, \bar{z}\,c).
\]
To ensure the various terms in the terminal condition are of similar order of magnitude, we choose $\beta = 1/\bar{z}\bar{S}$ and $\alpha = \alpha'/\bar{z}^2$.


Using the explicit form of $f$ and suppressing the variables that remain unaffected in the difference, we find the following PIDE for $\tilde{v}$
\begin{equation}
\label{eqn:PIDE-vtilde}
\begin{split}
\tfrac{1}{T} \partial_s \tilde{v} &+ \kappa^+\!\big(\bar{S}\,(\tfrac{b}{a} - x)\big)\;\Big[\tilde{v}\big(x+\tfrac{\theta^+}{\bar{S}}\big) - \tilde{v}\Big] 
\\
&+  \kappa^-\!\big(\bar{S}\,(\tfrac{b}{a} - x)\big)\;\Big[\tilde{v}\big(S-\tfrac{\theta^-}{\bar{S}}\big) - \tilde{v}\Big] 
\\
&
+ \lambda^x\!\big(\bar{S}\,(\tfrac{b}{a} - x)\big)\;\Big[\tilde{v}\Big(a + \tfrac{\pi^x}{\bar{r}^x}, b - f\big(a,b,\tfrac{\pi^x}{\bar{r}^x}\big)\Big) - \tilde{v}\Big] 
\\
&+  \lambda^y\!\big(\bar{S}\,(\tfrac{b}{a} - x)\big)\;\Big[\tilde{v}\Big(a - f\big(b,a,\tfrac{\pi^y}{\bar{r}^y}\big), b + \tfrac{\pi^y}{\bar{r}^y}\Big) - \tilde{v}\Big]
\\
&+\tfrac{1}{2\varphi}\left(\tilde{f}\big(a,b,\tfrac{\zeta}{\bar{z}^x}\big) + \tilde{v}\Big(a + \tfrac{\zeta}{\bar{r}^x}, b - f\big(a,b,\tfrac{\zeta}{\bar{r}^x}\big), c - \tfrac{\zeta}{\bar{z}^x}\Big) - \tilde{v}\right)_+^2= 0,
\end{split}
\end{equation}
with $\tilde{v}(T,x,a,b,c) = \tilde{f}(a,b,c) - \alpha'\,c^2$ and
$$\tilde{f}(a,b,\pi) = b\,\pi (1 - \phi)\left(a + \tfrac{\bar{z}^x}{\bar{r}^x}\;\pi(1-\phi)\right)^{-1}.
$$

The DGM approximates the function $\tilde{v}(s,x,a,b,c)$ with a neural network using the DGM architecture proposed in \cite{sirignano2018dgm}; see Appendix \ref{app:DGM} for a graphical description of the architecture. The approximation is denoted here by $u(s,x,a,b,c;\psi)$, where $\psi$ represents the DGM parameters. This neural network architecture consists of a feed forward layer, followed by $k$ Highway-like layers,
and finally another feed forward layer. In particular, we choose $k=3$ and each layer with 64 neurons. All activation functions, except for the output layer which has no activation, are the hyperbolic tangent. The loss function associated with the function approximation and the PIDE \eqref{eqn:PIDE-vtilde} is
\[
\mathcal{L}[\psi] = \frac{1}{B}\sum_{i=1}^B (\mathcal{D} u(\mathbf{x}_i;\psi))^2 + \frac{1}{B}\sum_{i=1}^B (u(\mathbf{y}_i;\psi) - \tilde{v}(\mathbf{y}_i))^2\;,
\]
where $\mathcal{D}$ denotes the PIDE given in \eqref{eqn:PIDE-v}, $(\mathbf{x}_i)_{i=1,\dots,B}  = (s^i,x^i,a^i,b^i, c^i)_{i=1,\dots,B}$ represents a minibatch sample of data in the region $[0,1]^2 \times [0.01, 1] \times [0,1]^2$, corresponding to random points in the interior of the transformed state space, and $(\mathbf{y}_i)_{i=1,\dots,B} = (1,x^i,a^i,b^i, c^i)_{i=1,\dots,B}$ is a minibatch sample independent of $(\mathbf{x}_i)_{i=1,\dots,B}$, corresponding to random points at terminal time but in the interior of the remaining transformed state space. 

We train the DGM neural network (using the architecture provided in Figures \ref{fig:networkoutside} and \ref{fig:dgmlayer3}, we refer to \cite{al2022extensions} for precise details)  by minimising the loss function using minibatch samples of $B=1,024$ in each iteration, for a total of 50,000 iterations, with  Adam update steps, using the learning rate scheme proposed in the original DGM article that goes from $10^{-4}$ to $10^{-7}$ with piece-wise constant decay.

We chose the following normalising constants $\bar{S} = 2\,S_0$, $\bar{r} = 2\,r^x_0$ and $\bar{z} = z^x_0$. After training, we obtain an approximation of $v$ in the domain $[0,T] \times [0, \bar{S}] \times [0.01 \, \bar{r}, \bar{r}] \times [0, \bar{S}\, \bar{r}] \times [0,\bar{z}]$. The lower bound for $r^x$ is necessary to avoid instability in the loss $\mathcal{L}[\psi]$. As this state variable corresponds to the number of token-$x$ in the pool, the choice of this lower bound has no material effect in practice.

\subsection{Results}

\begin{wraptable}{r}{0.5\textwidth}
\footnotesize
    \centering
    \begin{tabular}[t]{lr}
    \toprule\toprule
    Param. & Value \\
    \hline
    $S_0$ & 1,300 \\
    $\theta^{\pm}$ & 0.02\\
    $r^y_0$ & $5\times 10^7$\\
    $r^x_0$ & $r^x_0/S_0$\\
    $\pi_x$ & 1 \\
    $\pi_y$ & $\pi_x \times S_0$ \\
    \bottomrule
    \bottomrule
    \end{tabular}
    \hspace*{1em}
        \begin{tabular}[t]{lr}
        \toprule\toprule
    Param. & Value \\
    \hline
    $\phi$ & 0.3\%\\
    $z^x_0$ & 40 \\
    $z^y_0$ & 0 \\
    $\zeta$ & 2\\
    $\varphi$ & 2 \\
    $\alpha$ & 100 \\
    \bottomrule
    \bottomrule
    \end{tabular}
    \vspace*{1em}
    \caption{Parameter specifications for the numerical experiments.}
    \label{tab:exo_param}
\end{wraptable}
Drawing inspiration from the empirical data, we employ the parameter set listed in Table \ref{tab:exo_param} for our numerical experiments. 
After approximating the value function $\tilde{v}$ using DGM training, we transform the approximation to $v$ and use \eqref{eq:optimal_control} to approximate the optimal intensity process in feedback form.

\begin{figure}[H]
    \centering
    \begin{minipage}[t]{0.48\textwidth}
    \centering
    \vspace{0pt}
    \includegraphics[width=0.8\textwidth]{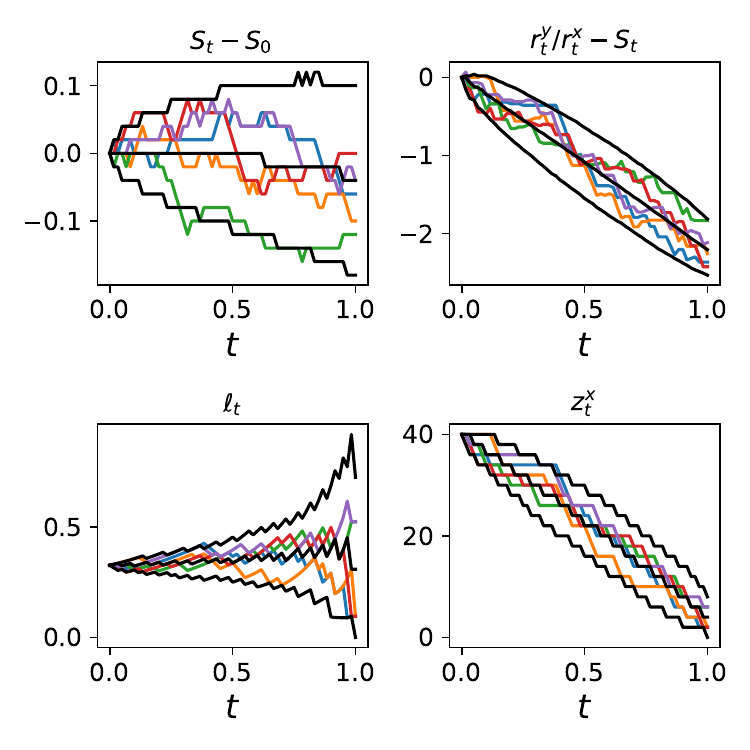}
    \vspace*{-1.5em}
    \caption{Five sample paths of $S_t$, the spread $r_t^y/r_t^x - S_t$, the optimal trading rate and the agent's inventory on crypto $x$ together with the $0.05$, $0.5$, and $0.95$ quantiles. 
    }
    \label{fig:sim_DGM}
    \end{minipage}
    \quad
    \begin{minipage}[t]{0.48\textwidth}
    \centering
    \vspace{0pt}
    \includegraphics[width=0.8\textwidth]{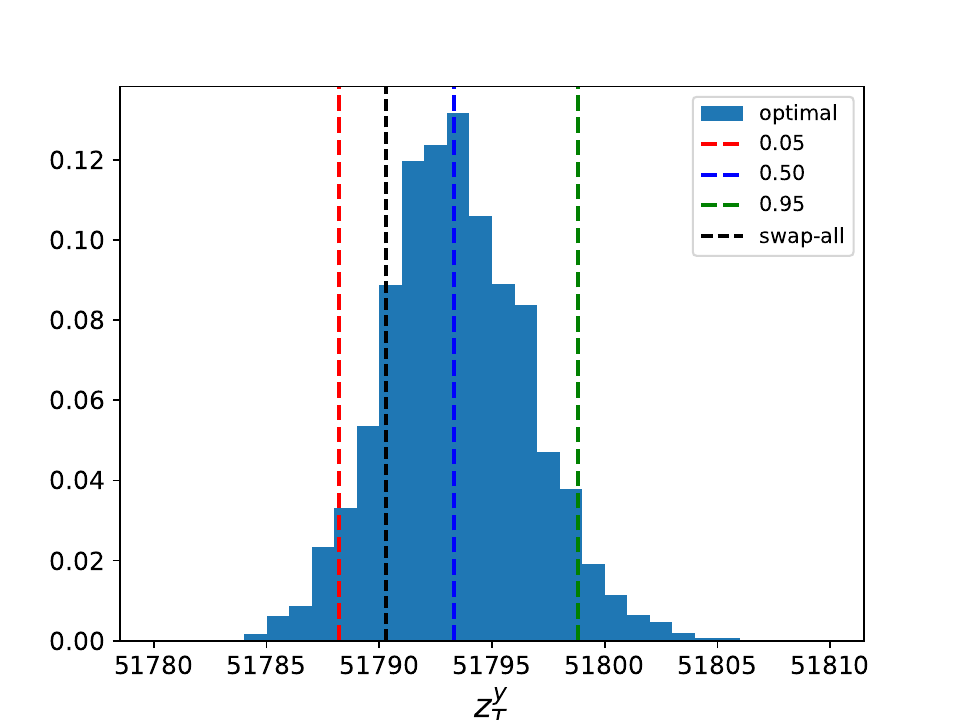}
    \caption{Histogram of terminal amount of token-$y$ received. The amount received if all were swapped at time $0$ is shown by the dotted line.}    
    \label{fig:zy_T_hist_DGM}
    \end{minipage}
\end{figure}

Figure \ref{fig:sim_DGM} portrays a few sample paths of the optimal strategy together with quantiles across all sample paths at each point in time. The bottom-left panel of the figures suggests that, when no orders are executed, the intensity at which the agent executes orders increases as time marches forward. Whenever an execution occurs, however, the intensity drops. The trading actions of the agent induces a spread 
between the marginal price and the centralised exchange price, as demonstrated in the top-right panel. There is an analogous small, but non-zero, downward impact in the centralised exchange price as seen in the top-left panel. The bottom-right panel shows that there tends to be some remaining token-$x$ to be liquidated at the end of the trading horizon.

In Figure \ref{fig:zy_T_hist_DGM}, we show the histogram of the final amount of token-$y$ the agent receives. The dashed-line indicates the amount of crypto-$y$ the agent receives if they swap all of token-$x$  at the start of the trading horizon. The figure illustrates that the  the optimal strategy provides an edge over full liquidation and hence avoids price slippage, moreover, the strategy beats the na\"ive strategy approximately $84.8\%$ of time. Table \ref{tab:zy-quantiles} shows the quantiles of $z_T^y$ resulting from the optimal strategy together with the   na\"ive strategy value.

\begin{table}[htbp]
  \centering
  \caption{Quantiles of the terminal amount of token-$y$ of the optimal strategy compared with the na\"ive strategy.}
    \begin{tabular}{rrrrr}
    \toprule
    \toprule
    $\alpha$ & 5\% & 50\% & 95\% & na\"ive
    \\
    \midrule
    $q_\alpha$ &  51,788.48 & 51,793.32 & 51,799.09 
    & 51,790.29
    \\
    \bottomrule\bottomrule
    \end{tabular}%
  \label{tab:zy-quantiles}%
\end{table}%

\begin{figure}
    \centering
\includegraphics[width=0.95\textwidth]{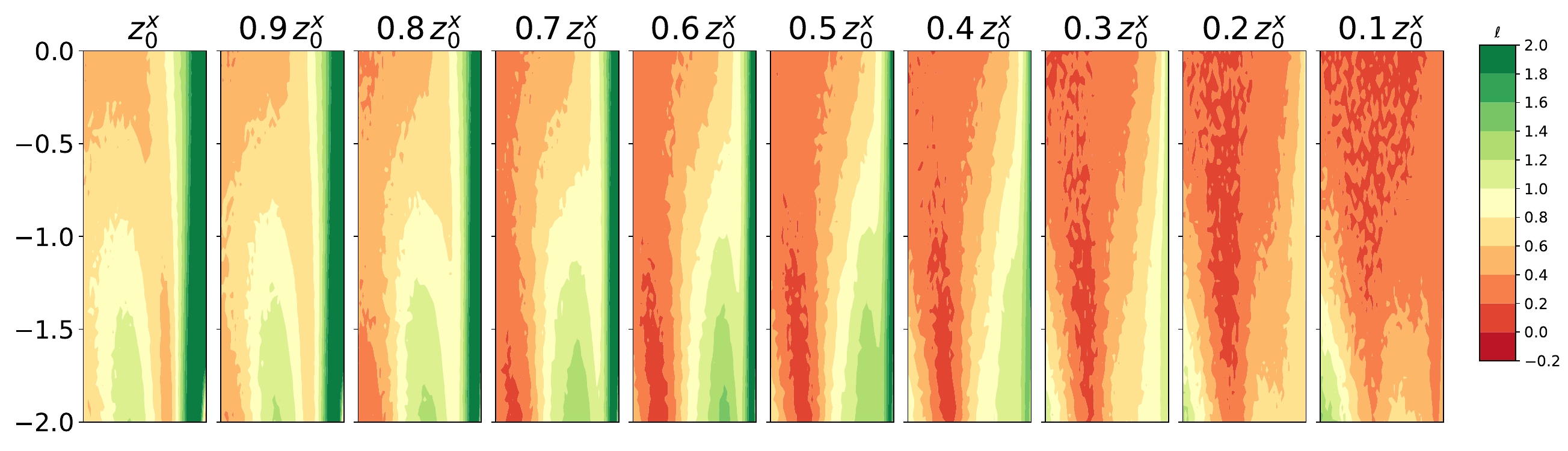}
    \caption{Optimal intensity of trades computed using the DGM approach varying in the vertical axis the spread, in the horizontal axis time from 0 to $T$ and each subplot denotes an initial inventory denoted on the top of the plot.}
    \label{fig:strategy_DGM}
\end{figure}

Finally, Figure \ref{fig:strategy_DGM} presents the optimal strategy as a function of three variables: (i) time on the x-axis, (ii) the spread $\frac{r_t^y}{r_t^x} - S_t$ on the y-axis, and (iii) the initial inventory of token-$x$ across panels. It shows that the optimal intensity usually increases as time marches forward, reduces as the initial inventory decreases, and generally increases as the spread widens.

    
\section{Conclusions}

We developed a model of how AMM swap events and centralised exchange prices co-move using detailed information about trades. Importantly, the model does not approximate processes as continuous diffusion, but rather models the microscopic aspects of the events and how they cross excite one another. Ultimately, prices are pure jump processes with stochastic intensities that all affect one another. Furthermore, we account for how trading actions of an agent modifies the market dynamics and then pose an stochastic control problem where the agent controls the intensity at which they send orders to the market. They do so to hide their trading actions. The agent wishes to maximise the total amount of token-$y$ they receive from liquidating all of token-$x$, but does so in a manner that accounts for how their trading actions alter the market. The resulting HJB equation is of moderate dimension, and not analytically tractable. Hence, we employ DGM to solve the HJB equation and obtain the optimal trading actions. We demonstrate that the trader does indeed achieve better performance than na\"ive liquidation of the asset, and avoids price slippage.

The framework we develop here can be used to solve other related optimisation problems, e.g.,  where the trader also sends orders to the centralised exchange, and where the trader is allowed to swap in and out of positions in both tokens. Furthermore, while we take a DGM approach to solving the resulting HJB equation, one could utilise reinforcement learning tools, such as double deep Q-learning, deep deterministic policy gradient, actor-critic, and so on, to directly solve the optimisation problem. As well, rather than optimising expectation of the cumulative amount of tokens the trader receives, we could apply dynamic risk measures to incorporate risk. Such criterion can be optimised using the methods developed in \cite{coache2021reinforcement} and \cite{coache2022conditionally}.


\appendix

\section{Additional Empirical Results}

\begin{figure}[!htbp]%
    \centering
    \includegraphics[width=0.7\textwidth]{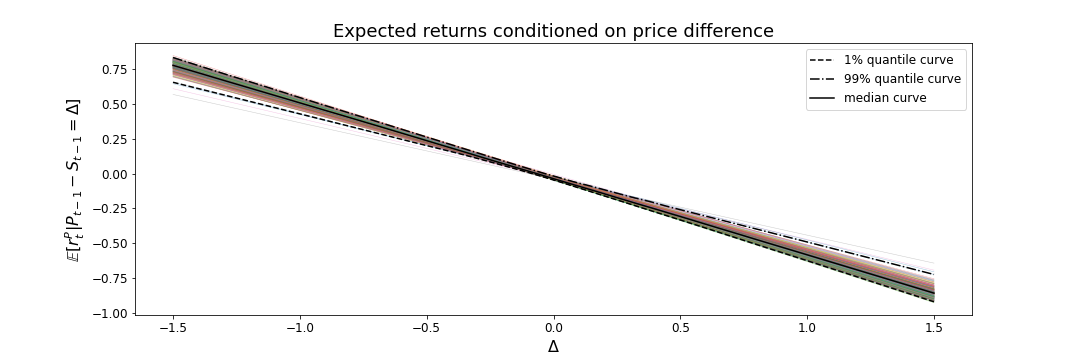}%
    \\
    \includegraphics[width=0.7\textwidth]{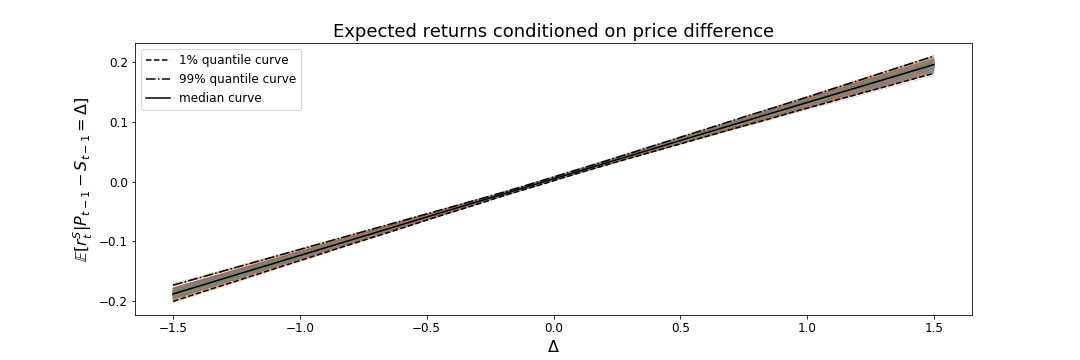} %
    \caption{Repeated estimates of the  expected value of returns conditioned on spread.}%
    \label{fig:expec-rets-cond-diff-many-regs}%
\end{figure}

\begin{figure}[!htbp]%
    \centering
    \includegraphics[width=0.7\textwidth]{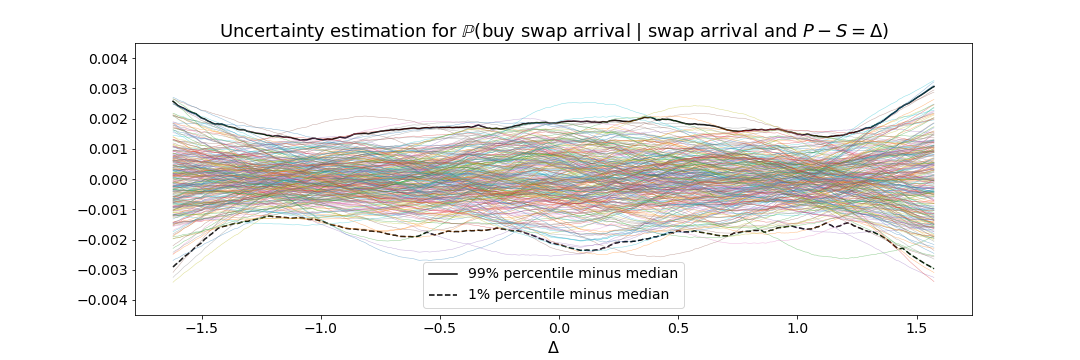}%
    \\
    \includegraphics[width=0.7\textwidth]{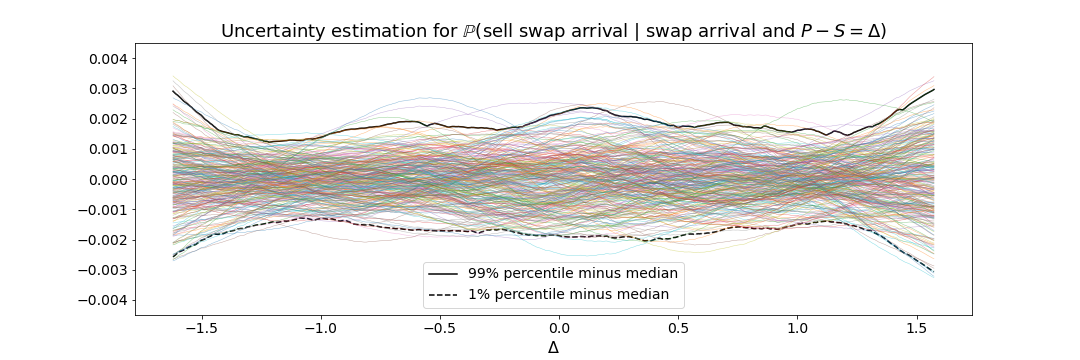}%
    \caption{Assessing the uncertainty in the estimation of Figure \ref{fig:swaps-probs-cond-diff}}%
    \label{fig:swap-probs-cond-diff-uncertainty}%
\end{figure}

\begin{figure}[!htbp]%
    \centering
    \includegraphics[width=0.7\textwidth]{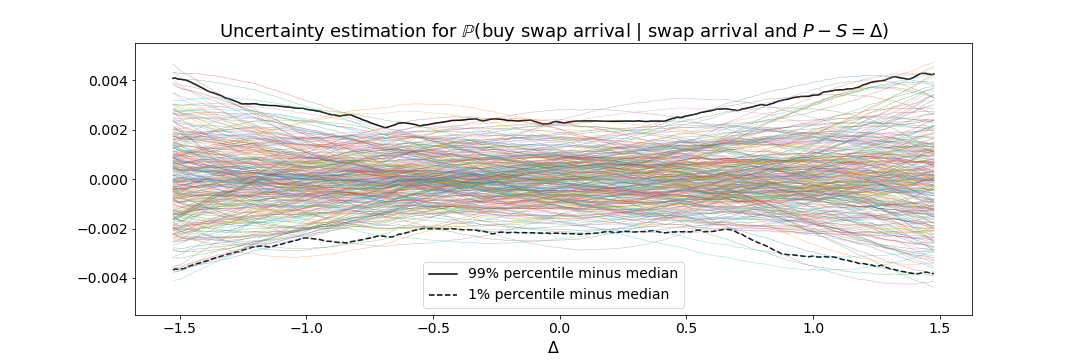}%
    \\
    \includegraphics[width=0.7\textwidth]{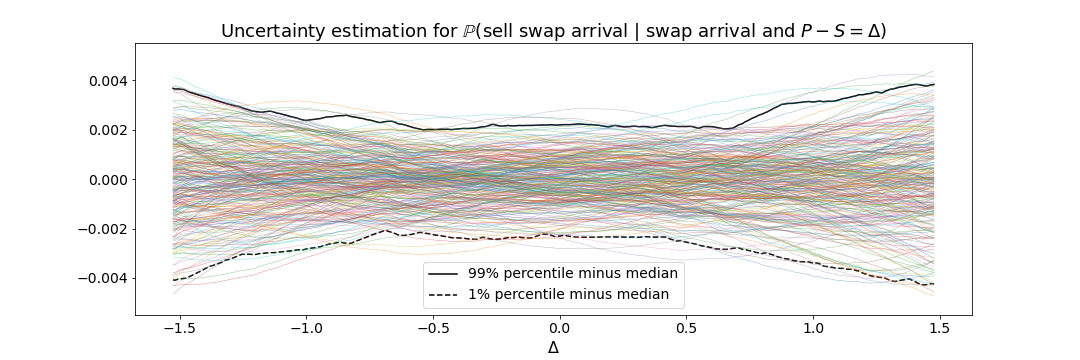}%
    \caption{Assessing the uncertainty in the estimation of Figure \ref{fig:trade-probs-cond-diff}.}%
    \label{fig:trade-probs-cond-diff-uncertainty}%
\end{figure}

\clearpage

\section{DGM Architecture}\label{app:DGM}

The following diagrams provide the details of the DGM architecture we use and is a reproduction of the one used in \cite{al2022extensions}.
\begin{figure}[h!]
	\centering
	\scalebox{0.7}{
	\begin{tikzpicture}
	
	\definecolor{yellowd}{RGB}{255,187,51}
	\definecolor{yellowf}{RGB}{255,221,153}
	\definecolor{greend}{RGB}{102,255,102}
	\definecolor{greenf}{RGB}{153,255,153}
	\definecolor{blued}{RGB}{51,187,255}
	\definecolor{bluef}{RGB}{153,221,255}
	\definecolor{pinkd}{RGB}{210,121,164}
	\definecolor{pinkf}{RGB}{230,179,204}
	
	\tikzstyle{scircle}=[circle, thick, draw=greend, fill=greenf, minimum size=1cm]
	\tikzstyle{xcircle}=[circle, thick, draw=blued, fill=bluef, minimum size=1cm]
	\tikzstyle{ycircle}=[circle, thick, draw=pinkd, fill=pinkf, minimum size=1cm]
	\tikzstyle{smallrect}=[rectangle, thick, draw=yellowd, fill=yellowf, minimum height=0.75cm, minimum width=1.5cm,]
	\tikzstyle{dgmlayer}=[rectangle, thick, draw=yellowd, fill=yellowf, minimum height=3cm, minimum width=1.5cm, rounded corners]
	\tikzstyle{arrow}=[draw=black, -{Latex[length=2.5mm]}, thick]

	\node[smallrect] (R1) {$\vw^1 \cdot \vx + \vb^1$};
	\node[scircle, below=1.0cm of R1] (S1) {$S^1$};
	\node[xcircle, above=1.0cm of R1] (X) {$\vx$};
	
	\node[dgmlayer, right=1.5cm of S1, yshift=1cm] (L1) {};
	\node[rotate=-90] at (L1) {DGM Layer};
	\node[dgmlayer, right=1.0cm of L1] (L2) {};
	\node[rotate=-90] at (L2) {DGM Layer};
	\node[dgmlayer, right=1.0cm of L2] (L3) {};
	\node[rotate=-90] at (L3) {DGM Layer};
	
	\node[scircle, right=1.0cm of L3, yshift=-1cm)] (SL) {$S^{L+1}$};
	\node[smallrect, right=of SL] (RL) {$\vw \cdot S^{L+1} + \vb$};
	\node[ycircle, right=of RL] (Y) {$\vy$};
	
	\draw[arrow] (X) -- (R1);
	\draw[arrow] (R1) -- node[anchor=east] {$\sigma$} (S1);
	\draw[arrow] (S1) -- (S1-|L1.west);
	\draw[arrow] (S1-|L1.east) -- (S1-|L2.west);
	\draw[arrow] (S1-|L2.east) -- (S1-|L3.west);
	
	\draw[arrow] (X) -| coordinate (aL1) (L1);
	\draw[arrow] (aL1) -| coordinate (aL2) (L2);
	\draw[arrow] (aL2) -| (L3);
	
	\draw[arrow] (S1-|L3.east) -- (SL);
	\draw[arrow] (SL) -- (RL);
	\draw[arrow] (RL) -- (Y);

	\end{tikzpicture}}
	\caption{The main components of the DGM architecture.}
	\label{fig:networkoutside}
\end{figure}
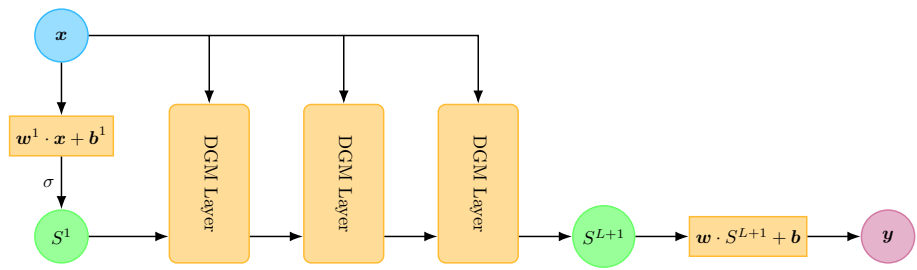

\begin{figure}[h!]
	\centering
	\scalebox{0.7}{
	\begin{tikzpicture}
	
	\definecolor{yellowd}{RGB}{255,187,51}
	\definecolor{yellowf}{RGB}{255,221,153}
	\definecolor{greend}{RGB}{102,255,102}
	\definecolor{greenf}{RGB}{153,255,153}
	\definecolor{blued}{RGB}{51,187,255}
	\definecolor{bluef}{RGB}{153,221,255}
	\definecolor{pinkd}{RGB}{210,121,164}
	\definecolor{pinkf}{RGB}{230,179,204}
	
	\tikzstyle{scircle}=[circle, thick, draw=greend, fill=greenf, minimum size=1cm]
	\tikzstyle{xcircle}=[circle, thick, draw=blued, fill=bluef, minimum size=1cm]
	\tikzstyle{ycircle}=[circle, thick, draw=pinkd, fill=pinkf, minimum size=1cm]
	\tikzstyle{smallrect}=[rectangle, thick, draw=yellowd, fill=yellowf, minimum height=1.5cm, minimum width=4.0cm, rounded corners]
	\tikzstyle{arrow}=[draw=black, -{Latex[length=2.5mm]}, thick]
	
	\node[] (SX1) {};
	\node[right=of SX1] (SX2) {};
	
	\node[xcircle, left=of SX1, yshift=1cm] (Sold) {$S$};
	\node[xcircle, left=of SX1, yshift=-1cm] (X) {$\vx$};
	
	\node[smallrect, right=of SX2] (GR) {$\vu^z \cdot \vx + \vw^z \cdot S + \vb^z$};
	\node[smallrect, above=of GR] (ZR) {$\vu^g \cdot \vx + \vw^g \cdot S + \vb^g$};
	\node[smallrect, below=of GR] (RR) {$\vu^r \cdot \vx + \vw^r \cdot S + \vb^h$};
	
	\node[scircle, right=of ZR] (Z) {$Z$};
	\node[scircle, right=of GR] (G) {$G$};
	\node[scircle, right=of RR] (R) {$R$};
	
	\node[smallrect, right= 1.45cm of Z] (SR) {$(1-G) \odot H + Z \odot S$};
	\node[smallrect, right=of R] (HR) {$\vu^h \cdot \vx + \vw^h \cdot (S \odot R) + \vb^h$};
	\node[scircle] (H) at ($(HR)!0.5!(SR)$) {$H$};
	\node[ycircle, right=of SR] (Snew) {$S^{new}$};

	\draw[thick] (Sold) -| (SX1.center);
	\draw[thick] (X) -| (SX1.center);
	\draw[arrow] (SX1.center) -- (GR);
	\draw[arrow] (SX2.center) |- (ZR);
	\draw[arrow] (SX2.center) |- coordinate (lRR) (RR);
	\draw[arrow] (ZR) -- node[above] {$\sigma$} (Z);
	\draw[arrow] (GR) -- node[above] {$\sigma$} (G);
	\draw[arrow] (RR) -- node[above] {$\sigma$} (R);
	\draw[arrow] (Z) -- (SR);
	\draw[arrow] (R) -- (HR);
	\draw[arrow] (HR) -- node[left] {$\sigma$} (H);
	\draw[arrow] (H) -- (SR);
	\draw[arrow] (G) -| ([xshift=-1.0cm]SR.south);
	\draw[arrow] (SR) -- (Snew);
	
	\node[above=of ZR] (aZR) {};
	\draw[thick] (Sold) |- (aZR.center);
	\draw[arrow] (aZR.center) -| (SR);
	\node[below=of RR] (bRR) {};
	\draw[thick] (lRR) |- (bRR.center);
	\draw[arrow] (bRR.center) -| (HR);

	\end{tikzpicture}}
	\caption{The operations within a single DGM layer.  Here $\odot$ denotes Hadamard (element-wise) multiplication, $\sigma$ is an activation function and the $\vu$, $\vw$ and $\vb$ terms with various superscripts are the model parameters.}
	\label{fig:dgmlayer3}
\end{figure}
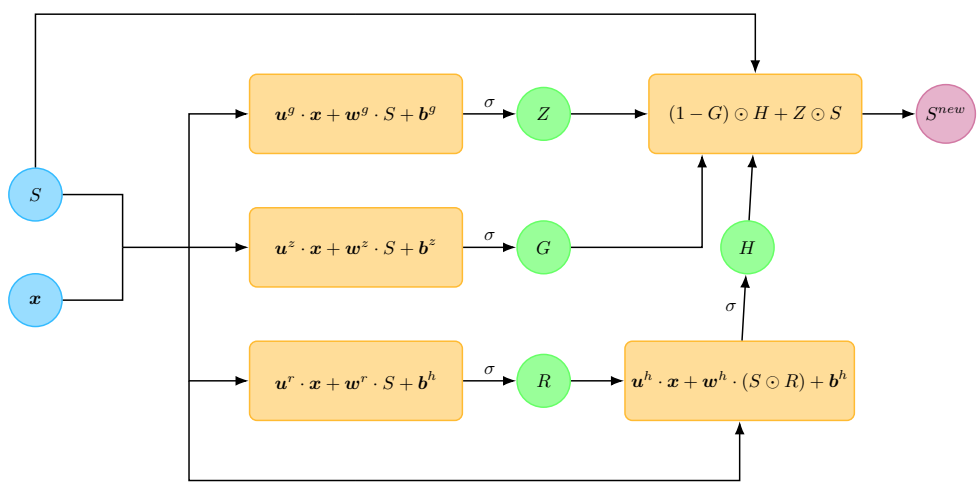


\begin{thebibliography}{10}

\bibitem{adams2020uniswap}
{\sc H.~Adams, N.~Zinsmeister, and D.~Robinson}, {\em Uniswap v2 core, 2020},
  URL: https://uniswap. org/whitepaper. pdf,  (2020).

\bibitem{adams2021uniswap}
{\sc H.~Adams, N.~Zinsmeister, M.~Salem, R.~Keefer, and D.~Robinson}, {\em
  Uniswap v3 core}, Tech. rep., Uniswap, Tech. Rep.,  (2021).

\bibitem{al2022extensions}
{\sc A.~Al-Aradi, A.~Correia, G.~Jardim, D.~de~Freitas~Naiff, and Y.~Saporito},
  {\em Extensions of the deep {G}alerkin method}, Applied Mathematics and
  Computation, 430 (2022), p.~127287.

\bibitem{al2018solving}
{\sc A.~Al-Aradi, A.~Correia, D.~Naiff, G.~Jardim, and Y.~Saporito}, {\em
  Solving nonlinear and high-dimensional partial differential equations via
  deep learning}, Report for the Financial Mathematics Team Challenge FTMC
  Brazil, 2018, available at arXiv:1811.08782,  (2018).

\bibitem{angeris2022constant}
{\sc G.~Angeris, A.~Agrawal, A.~Evans, T.~Chitra, and S.~Boyd}, {\em Constant
  function market makers: Multi-asset trades via convex optimization}, in
  Handbook on Blockchain, Springer, 2022, pp.~415--444.

\bibitem{angeris2022optimal}
{\sc G.~Angeris, A.~Evans, T.~Chitra, and S.~Boyd}, {\em Optimal routing for
  constant function market makers}, in Proceedings of the 23rd ACM Conference
  on Economics and Computation, 2022, pp.~115--128.

\bibitem{bergault2022automated}
{\sc P.~Bergault, L.~Bertucci, D.~Bouba, and O.~Gu{\'e}ant}, {\em Automated
  market makers: Mean-variance analysis of {LPs} payoffs and design of pricing
  functions}, arXiv preprint arXiv:2212.00336,  (2022).

\bibitem{cartea2022decentralised-es}
{\sc {\'A}.~Cartea, F.~Drissi, and M.~Monga}, {\em Decentralised finance and
  automated market making: Execution and speculation}, Available at SSRN,
  (2022).

\bibitem{cartea2022decentralisedl-llp}
{\sc {\'A}.~Cartea, F.~Drissi, and M.~Monga}, {\em Decentralised finance and
  automated market making: Predictable loss and optimal liquidity provision},
  Available at SSRN 4273989,  (2022).

\bibitem{cartea2023execution}
{\sc {\'A}.~Cartea, F.~Drissi, and M.~Monga}, {\em Execution and statistical
  arbitrage with signals in multiple automated market makers}, Available at
  SSRN,  (2023).

\bibitem{coache2021reinforcement}
{\sc A.~Coache and S.~Jaimungal}, {\em Reinforcement learning with dynamic
  convex risk measures}, arXiv preprint arXiv:2112.13414,  (2021).

\bibitem{coache2022conditionally}
{\sc A.~Coache, S.~Jaimungal, and {\'A}.~Cartea}, {\em Conditionally elicitable
  dynamic risk measures for deep reinforcement learning}, arXiv preprint
  arXiv:2206.14666,  (2022).

\bibitem{fan2022differential}
{\sc Z.~Fan, F.~J. Marmolejo-Coss{\'\i}o, B.~Altschuler, H.~Sun, X.~Wang, and
  D.~Parkes}, {\em Differential liquidity provision in uniswap v3 and
  implications for contract design}, in Proceedings of the Third ACM
  International Conference on AI in Finance, 2022, pp.~9--17.

\bibitem{fissler2016higher}
{\sc T.~Fissler and J.~F. Ziegel}, {\em Higher order elicitability and
  osband’s principle},  (2016).

\bibitem{neuder2021strategic}
{\sc M.~Neuder, R.~Rao, D.~J. Moroz, and D.~C. Parkes}, {\em Strategic
  liquidity provision in uniswap v3}, arXiv preprint arXiv:2106.12033,  (2021).

\bibitem{oksendal2007applied}
{\sc B.~K. {\O}ksendal and A.~Sulem}, {\em Applied stochastic control of jump
  diffusions}, vol.~498, Springer, 2007.

\bibitem{sirignano2018dgm}
{\sc J.~Sirignano and K.~Spiliopoulos}, {\em {DGM}: A deep learning algorithm
  for solving partial differential equations}, Journal of computational
  physics, 375 (2018), pp.~1339--1364.

\end{thebibliography}
\end{document}